\documentstyle[prl,aps,multicol,epsf,graphics]{revtex}
\def\d{{\rm d}}
\begin{document}
\draft
%\preprint{IASGZ 75/96/}
%\preprint{cond-mat/}
\title{Statistical mechanics of double-stranded semi-flexible polymers}
\author{Tanniemola~B. Liverpool$^{\dag}$, Ramin Golestanian$^{\S}$ and Kurt Kremer$^{\dag}$}
\address{$^{\dag}$Max-Planck-Institut f{\"u}r Polymerforschung,
D-55021 Mainz, Germany
\\ $^{\S}$Institute for Advanced Studies in Basic Sciences,
Zanjan 45195-159, Iran}
\date{\today}
\maketitle
\begin{abstract}
  We study the statistical mechanics of double-stranded semi-flexible polymers
  using both analytical techniques and simulation.  We find a transition at
  some finite temperature, from a type of short range order to a fundamentally
  different sort of short range order. In the high temperature regime, the
  2-point correlation functions of the object are identical to worm-like
  chains, while at low temperatures they are different due to a
  twist structure. In the low temperature phase, the polymers develop a
  kink-rod structure which could clarify some recent puzzling experiments on
  actin.
\end{abstract}
\pacs{87.15By, 36.20Ey, 61.25Hq}
\begin{multicols}{2}

  A double-stranded semi-flexible polymer chain is the basic structure of many
  biopolymers of which examples are DNA and proteins such as actin. There has
  been a lot of recent experimental interest in systems of biopolymers ranging
  from the elasticty of biopolymer networks and its use in the prediction of
  the mechanical properties of cells to the direct visualisation of single
  chain properties. The model most used in the study of biopolymers is that of
  the worm-like chain~\cite{KratPor} in which the polymer flexibility
  (structure) is determined by a single length, the persistence length
  $\ell_{p}$ which measures the tangent--tangent correlations.  For example,
  DNA has a persistence length $\ell_{p} \approx 50\mbox{nm}$ whilst for actin
  $\ell_{p} \approx 17\mu\mbox{m}$. These biopolymers are known to have a more
  complex `twisted' structure.  The multi-stranded nature of these polymers is
  also not taken into account in a simple worm-like chain model.  It is not
  clear if such a fine structure will have an effect on the global properties
  of these objects. A possible effect of such fine structure is what we
  attempt to study in this Letter. Our model is, in a sense, microscopic
  because the interaction between the bend and twist degrees of freedom is a
  {\it result}. This is fundamentally diffent from previous
  approaches~\cite{Marko} which try to include the twist degrees of freedom by
  adding extra terms to the free energy.

We study a version of the rail-way track model of
Everaers-Bundschuh-Kremer (EBK) \cite{EBK} for a double-stranded
semi-flexible polymer, embedded in a $d$-dimensional space for arbitrary
$d$. Excluded volume and electrostatic interactions have been ignored. We find
that the system, has qualitatively different properties in the low temperature
and high temperature regimes, in contrast
to what one might naively expect from an inherently one-dimensional
system with local interactions and constraints. The tangent--tangent
correlation function decays exponentially in the whole range of
temperatures with a ``tangent-persistence length'' $\tilde{\ell}_{\rm TP}$
that has a very slow temperature
dependence, and whose scale is determined by the (bare) persistence
length of a single strand $\ell_p = \kappa/k_{B}T$ ($\kappa$ is
the bending stiffness of a single strand). Note that it is independent
of $a$, the separation of the two strands, which is the other relevant
length scale in the problem.
However, the correlation function of the bond-director field,
defined as a vector that determines the separation and coupling of the two
strands of the combined polymer system, has different behaviour below and
above the temperature $T_c \simeq 4.27 \kappa/d k_{B} a$.
While it decays purely exponentially for $T > T_c$, there are additional
oscillatory modulations for $T < T_c$. The related ``bond-persistence length''
$\tilde{\ell}_{\rm BP}$ does not change appreciably at high temperatures
, where its scale is again set by $\ell_p$ alone.
In the low temperature phase, however, $\tilde{\ell}_{\rm BP}$ does show
a temperature dependence. In particular,
$\tilde{\ell}_{\rm BP} \sim \ell_{p}^{1/3} a^{2/3} \propto T^{{-1/3}}$ for $T
\rightarrow 0$, while $\tilde{\ell}_{\rm BP} \sim \ell_{p}$ for $T \sim
T_c$. Similarly, the ``pitch'' $H$, defined as the period of oscillations in
the low temperature regime, changes drastically with temperature
, ranging from $H \sim \tilde{\ell}_{\rm BP} \sim \ell_{p}^{1/3} a^{2/3}$
near $T=0$, to $H \sim 0$ near $T=T_c$.
At $T=0$ we regain a flat ribbon which has true long-range order
in both the tangent and bond-director fields. The ribbon is essentially a
rigid rod. As we approach $T=0$, the persistence lengths
and the pitch diverge with the scaling
$H \sim \tilde{\ell}_{\rm BP} \sim \tilde{\ell}_{\rm TP}^{1/3}$.

The system is composed of two semi-flexible chains, each with rigidity
$\kappa$, whose embeddings in $d$-dimensional space are defined by ${\bf r}_1
(s)$ and ${\bf r}_2 (s)$.  The Hamiltonian of the system can be written as the
sum of the Hamiltonians of two worm-like chains subject to the inextensiblity
constraints~\cite{semiflex}. The ribbon structure is enforced by having ${\bf
  r}_2(s')$ separated from ${\bf r}_1(s)$ by a distance $a$, i.e. ${\bf
  r}_2(s') = {\bf r}_1(s)+ a{\bf n}(s)$ where $|s-s'|$ can be non-zero but is
small. We have defined a bond-director field ${\bf n}(s)$, a unit vector
perpendicular to {\it both} strands.  The chains are assumed to have
``permanent'' bonds (e.g. hydrogen bonds) that are strong enough to keep the
distance between the two strands constant. In Ref.\cite{EBK}, it is argued
that the relevant constraint on the system would then be that in a bent
configuration the arc-length mismatch between the two strands should be very
small. We can calculate the arc-length mismatch for the bent configuration as
$ \Delta s=|{\bf r}_2 (s)-{\bf r}_1 (s)+a \;{\bf n}(s)| $, where $a$ is the
separation of the strands. We impose the constraint as a hard one, namely, we
set $\Delta s =0$, as opposed to Ref.\cite{EBK}.  Physically this means we
{\it do not} allow bends in the plane of the ribbon.  These bends are less
important in $d>2$ because as we shall see the lower length-scale will be set
by the `pitch' which will make the in plane fluctuations of the ribbon
irrelevant~\cite{EBK}.  This simplifying assumption should not change the
behavior of the system~\cite{hard}.  We implement the constraint $\Delta s =0$
by introducing the ``midcurve'' ${\bf r}(s)$: $ {\bf r}_1 (s) = {\bf
  r}(s)+\frac{a}{2} \; {\bf n}, \, \, \, {\bf r}_2 (s) = {\bf
  r}(s)-\frac{a}{2} \; {\bf n}.  $ In terms of the tangent to the midcurve
${\bf t}=\d {\bf r}/\d s$ and the bond-director ${\bf n}$ the Hamiltonian of
the system can now be written as
\begin{equation}
{\cal H}=\frac{\kappa}{2} \; \int \d s \; \left[
2 \left(\frac{\d {\bf t} (s)}{\d s} \right)^2
+\frac{a^2}{2}\left(\frac{\d^2 {\bf n}(s)}{\d s^2} \right)^2
\right],\label{Htn}
\end{equation}
subject to the exact (local) constraints
\begin{eqnarray}
({\bf t} \pm \frac{a}{2} \;\frac{\d {\bf n}}{\d s})^2=1, \qquad
{\bf n}^2=1, \qquad 
({\bf t} \pm \frac{a}{2} \;\frac{\d {\bf n}}{\d s}) \cdot {\bf n}=0.
\label{constr}
\end{eqnarray}
This completes the formulation of the model.

The statistical mechanics of semi-flexible chains are difficult due to the
constraint of in-extensibility. Various approximation methods have been
devised to tackle the problem. A successful scheme, that somehow manages to
capture the crucial features of the problem, is to impose global (average)
constraints rather than local (exact) ones \cite{semiflex}. This approximation
is known to be good for calculating the average end-to-end length but not so
good for the whole distribution. It corresponds to a saddle-point evaluation
of the integrals over the Lagrange multipliers, that are introduced to
implement the constraints \cite{semiflex}. In this sense, it is known to be a
``mean-field'' approximation in spirit.  To study the effects of fluctuations
on the mean-field result, we have performed a $1/d$-expansion similar to the
one successfully used by David and Guitter to study the crumpling transition
of crystalline membranes \cite{David}. We see that no divergent behaviour
appears in the diagrams of the {\it 2-point} correlation
functions which means that the mean-field behavior of these functions is not
changed by fluctuations. This does not preclude differences in higher
order correlation functions.

With the above discussion as justification, we apply the same approximation
scheme to our problem defined above. The local constraints in
Eq.(\ref{constr}) are relaxed to global ones. To do this, we add the
corresponding ``mass terms'' to our Hamiltonian $ \frac{{\cal H}_{m}}{k_{B}T}
= \int \d s \left[ \frac{b}{\ell_p}({\bf t}-\frac{a}{2} \;\frac{\d {\bf n}}{\d
    s})^2 +\frac{b}{\ell_p}({\bf t}+\frac{a}{2} \;\frac{\d {\bf n}}{\d s})^2
  +\frac{c a^2}{4 \ell_{p}^3} {\bf n}^2 \right .$ $\left . +
  \frac{e}{\ell_p} ({\bf t}-\frac{a}{2} \;\frac{\d {\bf n}}{\d s}) \cdot {\bf
    n} +\frac{e}{\ell_p} ({\bf t}+\frac{a}{2} \;\frac{\d {\bf n}}{\d s}) \cdot
  {\bf n} \right],\nonumber $ where $b$, $c$, and $e$ are dimensionless
constants.
%The partition function is then given by $Z[{\bf J},{\bf K}]= \int [{\cal
%D}{\bf t}(s)][{\cal D}{\bf n}(s)] \exp\left({-\frac{{\cal H}+{\cal
%H}_{m}}{k_{B}T}+{\bf J}\cdot{\bf t}+{\bf K}\cdot{\bf    n} }\right)$ 
%
We then determine the constants self consistently by demanding the
constraints of Eq.(\ref{constr}) to hold on average, where the thermal average
is calculated by using the total Hamiltonian ${\cal H}+{\cal H}_m$.  Note that
in choosing the above form, we have implemented the ``label symmetry'' of the
chains, namely, that there is no difference between two chains. The
self-consistency lead to the following set of equations for the constants $b$
and $c$: $ \frac{1}{4 \sqrt{2 b}}+\frac{a^2 \sqrt{c}}{4 d
  \ell_{p}^2}=\frac{1}{d}, \, \, \, c (b+\sqrt{c}) = \frac{d^2 \ell_{p}^4}{2
  a^4}, \nonumber $ and $e=0$. The above equations, which are nonlinear and
difficult to solve exactly, determine the behavior of $b$ and $c$ as a
function of $u=a/\ell_p$. We have solved them numerically in $d=3$ and the
solutions are given in Fig.~1.  One can solve the equations analytically in
two limiting cases. For $u \ll 1$ we find $b=d^2/32$ , and
$c=(d/\sqrt{2})^{4/3} u^{-8/3}$, whereas for $u \gg 1$ we find $b=d^2 /8$, and
$c=4/u^4$. In Fig.~1, the behavior of $b$ and $c$ is plotted as a function of
$u$. Note that $u$ is proportional to $T$ and can be viewed as a measure of
temperature.

\begin{figure}
\epsfxsize 6cm \rotatebox{-90}{\epsffile{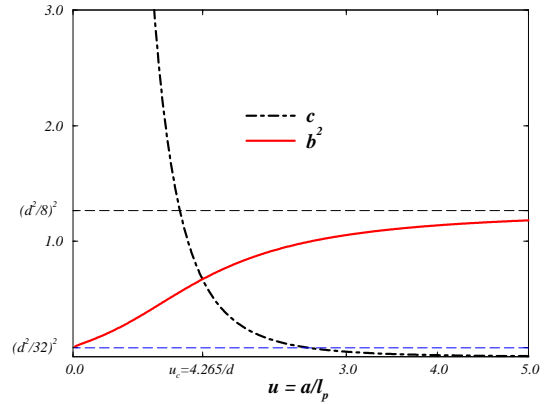}}
\protect\begin{minipage}[t]{8.5cm}{\centerline{\caption{The solution of the
        self consistent equations for the constants $b$ and $c$ as a function
        of $u = a/\ell_{p}$ in $d=3$. The value $u_{c} \simeq 4.27/d$
        corresponds to the transition point.}}}\end{minipage}
\label{fig1}
\end{figure}

We can then calculate the correlation functions.
For the tangent--tangent correlation one obtains
\begin{equation}
\left<{\bf t}(s) \cdot {\bf t}(0)\right>=\frac{d}{4 \sqrt{2 b}}
\exp\left(-\sqrt{2 b} \;\frac{s}{\ell_p}\right),\label{tt}
\end{equation}
whereas for the bond-director field one obtains
\end{multicols}
\begin{equation}
\left<{\bf n}(s) \cdot {\bf n}(0)\right>=
\frac{d \ell_{p}^2}{2 a^2 \sqrt{b^2-c}} \left(
\frac{\exp\left(-(b-\sqrt{b^2-c})^{1/2}\;\frac{s}{\ell_p}\right)}
{(b-\sqrt{b^2-c})^{1/2}}
-\frac{\exp\left(-(b+\sqrt{b^2-c})^{1/2}\;\frac{s}{\ell_p}\right)}
{(b+\sqrt{b^2-c})^{1/2}} \right).\label{nn}
\end{equation}
\begin{multicols}{2}
The tangent--tangent correlation (Eq.(\ref{tt})) is exactly what
we obtain for a single worm-like chain, and implies uniform
behavior for all temperatures. Eq.(\ref{nn}) on the other hand,
indicates a change of behavior at $b^2=c$ for the bond-director
correlation. The correlation is {\it overdamped} for $b^2 > c$
(high temperatures), while it is {\it underdamped} (oscillatory)
for $b^2 < c$ (low temperatures). The interesting point $b^2=c$
happens for $u_c=2^{19/4}/(2+\sqrt{2})^{3/2} d \simeq 4.27/d$, that
leads to the value for $T_c$ quoted above (see Fig.~1). 
We also find a divergence in the specific heat,
$C_{V}=\frac{\partial^{2}F}{\partial T^{2}}$ where $F=-k_{B}T\log Z$ at
$T_{c}$. It should be noted that it is not a thermodynamic phase transition in
the sense of long-range ordering and broken symmetry. It is a 
cross-over that appears due to competing effects, and the transition
is from a state with some short-range order to a state with a different
short-range order. Similar phenomena have been observed in Ising-like spin
system with competing interactions~\cite{Hornreich} and the
cross-over ({\it transition}) point corresponds to a type of {\it`Lifshitz
  point'} for a 1-d system. 

The nature of competition in our double-stranded polymer system can be
understood using a plaquette model. Consider rectangular plaquettes attached
to each other to form a ribbon. The attached sides (edges) correspond to the
bond-director field, while the other sides determine the tangent-director
field. The twist degree of freedom corresponds to twisting the plaquettes
against each other. The energy expression corresponding to bends comes from
the product of the tangent-directors of the neighbouring plaquettes, and has
no competition. On the other hand, the energy for twists that comes from the
product of the bond-director fields of the neighboring plaquettes, does show
competition. These competing effects come from an effective interaction
between bond-directors that are next nearest neighbors.  This is due to the
fact that alike twists meeting at an edge tend to unwind (annihilate) each
other, while unlike twists when they meet are trapped; they do not annihilate
each other.  This competition is only present at nonzero temperatures and is
merely due to topological constraints of the ribbon.

It is useful to study the bond-director correlation in the
limiting case $b^2 \ll c$, that corresponds to relatively low
temperatures. Using the asymptotic forms for $b$ and $c$, one obtains
$
\left<{\bf n}(s) \cdot {\bf n}(0)\right>=
\sqrt{2}\;\exp\left(-\left(\frac{d}{4 \ell_p a^2}\right)^{1/3} s\right)
\;\sin\left(\left(\frac{d}{4 \ell_p a^2}\right)^{1/3} s+\frac{\pi}{4}
\right),
$
for very low temperatures.
From the above expressions for the correlation functions, one can
read off the persistence lengths $\tilde{\ell}_{\rm TP}$ and
$\tilde{\ell}_{\rm BP}$, and the pitch $H$, as summarized above.

An intriguing feature of the behaviour of this model is that, although the
ground state ($T=0$) configuration of the system is a flat ribbon, and
supports no twists, upon raising the temperature, a twisted structure with
short range order emerges. We have confirmed this by performing extensive
Molecular Dynamics(MD)/Monte Carlo (MC) simulations of double-stranded
semi-flexible polymers. A bead-spring model with bending and stretching
energies was used. We combined a velocity-Verlet MD coupled to a heat bath
with an off-lattice pivot MC algorithm. The MD was useful for equilibriating
the shorter length-scales and MC the long length-scales. Details of the
algorithm and computational method will be presented elsewhere~\cite{inprep}.

\begin{figure}
\epsfxsize 6cm \rotatebox{-90}{\epsffile{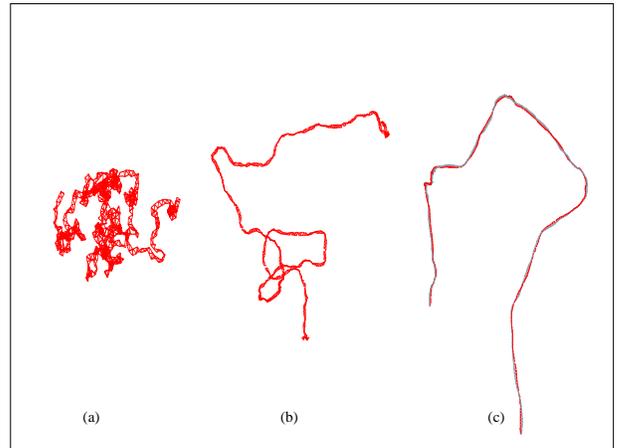}}
\protect\begin{minipage}[t]{8.5cm}{\centerline{\caption{Typical conformations from MD/MC simulations of a ribbon made up  of two chains of 400 monomers (a) above  (b) near and (c) below $T_{c}$}}}\end{minipage}
\label{fig2}
\end{figure}

We show typical equilibriated conformations in Fig.~2. The snapshots of the
polymer configurations suggest that at low temperatures the polymer can be
viewed as a collection of long, twisted (straight) rods that are connected by
short, highly curved sections of chain which we call ``kinks''. This picture
can be accounted for using a simple argument. We can model our system of two
semiflexible polymers subject to the constraint of constant separation, as a
semiflexible ribbon, i.e. a semiflexible linear object with anisotropic
rigidities whose Hamiltonian reads
\begin{equation}
{\cal H}= \frac{1}{2} \int \; \d s \; \sum_{{i,j}} \; \kappa_{ij} \;
\left(\frac{\d {\bf t}}{\d s}\right)_{i}\; 
        \left(\frac{\d {\bf t}}{\d s}\right)_{j}, \label{kij}
\end{equation} 
where $\kappa_{ij}=\kappa_{\parallel}\; n_i n_j+\kappa_{\perp}\;
(\delta_{ij}-t_i t_j-n_i n_j) $ determines the rigidity anisotropy of the
ribbon, corresponding to bending parallel or perpendicular to the
bond-director field. The ribbon structure would require $\kappa_{\parallel}
\gg \kappa_{\perp}$. To be consistent with the hard constraint (see above) of
constant separation of the polymers, we should take the limit of infinite
$\kappa_{\parallel}$. The Boltzman weight with the above Hamiltonian in the
$\kappa_{\parallel} \to \infty$ limit can be easily shown to require the
constraint $ \frac{\d {\bf t}(s)}{\d s} \cdot {\bf n}(s) = 0, \label{rodkink}
$ to hold exactly at {\it every} point of the ribbon. Recalling $\d {\bf
  t}/{\d s}= H(s) \;{\bf e}$ from the Frenet-Seret equations, where $H(s)$ is
the curvature at each point and $ {\bf e}$ is a unit normal vector to the
curve, we can write the constraint as $H(s) \; {\bf e}(s) \cdot {\bf n}(s)
=0$. It tells us that at each point either $H(s)=0$, that is we have straight,
twisted sections with no curvature, or ${\bf e}(s) \cdot {\bf n}(s) =0$,
curved regions where the the bond-director is perpendicular to the curve
normal. The former case would correspond to the rod segments whereas the
latter would correspond to the ``kinks''. Performing the configuration sum we
obtain that the average separation of the kinks is equal to the
tangent-persistence length, which is quite large at small enough temperatures.
As the temperature is raised, the kinks get closer so that at some temperature
the average kink separation becomes comparable to the size of the kinks, where
the rod-kink pattern disappears. This analysis can be understood in the
context of the mean-field $(\bf{n},\bf{t})$ model above by observing that at
low temperatures $\tilde{\ell}_{\rm BP} \ll \tilde{\ell}_{\rm TP}$, one can
imagine that there are roughly speaking rod-like segments of length
$\tilde{\ell}_{\rm TP}$, each supporting a number of shorter segments of
length $\tilde{\ell}_{\rm BP}$ that are twisted, but de-correlated with one
other. As the temperature is raised, the number of twisted rods in each
segment $N=\tilde{\ell}_{\rm TP}/\tilde{\ell}_{\rm BP}$ decreases very quickly
until it saturates to unity at $T=T_c$. For higher temperatures the mechanism
changes, and the bond correlations are cut off by the tangent fluctuations.
Hence, the short-range twist order does not exist anymore. All the main
features of the above picture have been observed in the simulation.  We plot
the $\langle {\bf n}(s) \cdot {\bf n}(0)\rangle$ correlation function from the
simulation in Fig.~3.
\begin{figure}
\epsfxsize 6cm \centerline{\epsffile{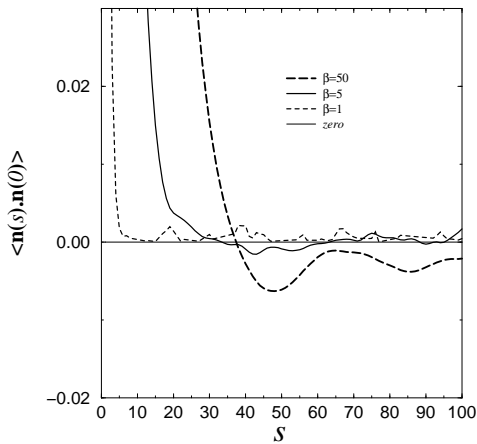}}
\protect\begin{minipage}[t]{8cm}{\centerline{\caption{The $\langle {\bf n}(s)
        \cdot {\bf n}(0)\rangle$ correlation function measured in the
        simulations for temperatures $\beta= 1/k_{B}T=1,5,50$ corresponding to
        $b^{2} >c,b^{2}\approx c \,\, \mbox{and} \,\, b^{2} < c$.  The
        averages were done over $\sim 10^{4}$ statistically independent
        samples.}}}\end{minipage}
\label{fig3}
\end{figure}

In conclusion, we have studied the properties of a well-defined model of a
double-stranded semi-flexible polymer using a mean-field analytical approach
as well as extensive MD/MC simulations.  We have shown novel non-trivial
differences between the high, low {\it and} zero temperature behaviour. An
alternative formulation of the problem (see Eq.(\ref{kij}) and the following
paragraph) where the constraints are implemented in a different way also
suggests qualitatively similar results~\cite{inprep}.  We also note that the
low temperature phase can be used to explain some recent experiments done on
single actin filaments by K\"as et al~\cite{Kas} who observed a length-scale
dependent bending rigidity {\it as well} as the kink-rod structure explained
above. The high-$q$ small bending rigidity can be associated with the kinks
and the low-$q$ large bending rigidity with the rods. Of course, upon
averaging the object would have worm-like chain statistics.  Finally we
mention that there are lots of interesting prospects for the study of such
complex linear objects. It is known that biopolymers, such as microtubules,
can also be multi-stranded objects. This method could be easily be extended to
describe triple-stranded objects.  The effect of an intrinsic twist changes
the ground state but does not change any of the conclusions of our
description~\cite{inprep} though we expect it to make the effective
persistence length much higher.

%\acknowledgments
We have benefited from numerous discussions with R. Bundschuh, R. Ejtehadi, R
Evaerers, E. Frey, G. Grest, M. Kardar and M. P\"utz. We would like to thank
the hospitality of ICTP, Trieste, where some of this work was done.

\end{multicols}

\begin{references}

\bibitem{KratPor}
O. Kratky and G. Porod, {\it Rec. Trav. Chim.}, {\bf 68}, 1106 (1949) .
 
\bibitem{Marko} J.F. Marko and E.D. Siggia, {\it Macromolecules}, {\bf 27},
  981 (1994); B. Fain, J. Rudnick and S. Ostlund, {\it preprint}
  condmat/9610126, (1996).

\bibitem{EBK} R. Everaers, R. Bundschuh, and K. Kremer, {\it Europhys. Lett.},
  {\bf 29}, 263 (1995).

\bibitem{semiflex} M.G. Bawendi and K.F. Freed, {\it J. Chem. Phys.}, {\bf
    83}, 2491 (1985); J.B. Lagowski, J. Noolandi, and B. Nickel, {\it J. Chem.
    Phys.}, {\bf 95}, 1266 (1991); A.M. Gupta, and S.F. Edwards, {\it J. Chem.
    Phys.}, {\bf 98}, 1588 (1993); T.B. Liverpool, and S.F. Edwards, {\it J.
    Chem. Phys.}, {\bf 103}, 6716 (1995).

\bibitem{hard}
In Ref.\cite{EBK} it is shown that, if we impose a soft
constraint using an energy term like $(k/2) \int (\Delta s)^2$,
we can see that the length $l=(\kappa/k a^2)^{1/2}$ determines two
different regimes; the interesting one being $L \gg l$ ($L$ is the length
of the chains). Hence, our hard constraint in fact, only
restricts us to the case of interest.

\bibitem{David}
 F. David, and E. Guitter, {\it Europhys. Lett.}, {\bf 5}, 709 (1988).

\bibitem{Hornreich}
R.M. Hornreich, R. Liebmann, H.G. Schuster and W Selke, {\it Z. Phys.}, {\bf
  B 35}, 91 (1979).

\bibitem{inprep}
T.B. Liverpool and K. Kremer, {\it in preparation}; R. Golestanian and T.B. Liverpool, {\it in preparation}.

%\bibitem{geometry}
%B.A. Dubrovin, A.T. Fomenko and S.P. Novikov, {\it Modern Geometry - Methods
%  and Applications Part I: The Geometry of Surfaces, Transformation groups and
%  Fields}, (GTM Springer-Verlag, New York, 1984). 

\bibitem{Kas}
J. K\"as, H. Strey, M. B\"armann and E. Sackmann, {\it Europhys. Lett.},
{\bf 21}, 865 (1993). 

\end{references}
\end{document}